# Decoding the Threat Landscape : ChatGPT, FraudGPT, and WormGPT in Social Engineering Attacks

**Polra Victor Falade**

Cyber Security Department, Nigerian Defence Academy, Afaka, Kaduna, Nigeria

**ABSTRACT**

In the ever-evolving realm of cybersecurity, the rise of generative AI models like ChatGPT, FraudGPT, and WormGPT has introduced both innovative solutions and unprecedented challenges. This research delves into the multifaceted applications of generative AI in social engineering attacks, offering insights into the evolving threat landscape using the blog mining technique. Generative AI models have revolutionized the field of cyberattacks, empowering malicious actors to craft convincing and personalized phishing lures, manipulate public opinion through deepfakes, and exploit human cognitive biases. These models, ChatGPT, FraudGPT, and WormGPT, have augmented existing threats and ushered in new dimensions of risk. From phishing campaigns that mimic trusted organizations to deepfake technology impersonating authoritative figures, we explore how generative AI amplifies the arsenal of cybercriminals. Furthermore, we shed light on the vulnerabilities that AI-driven social engineering exploits, including psychological manipulation, targeted phishing, and the crisis of authenticity. To counter these threats, we outline a range of strategies, including traditional security measures, AI-powered security solutions, and collaborative approaches in cybersecurity. We emphasize the importance of staying vigilant, fostering awareness, and strengthening regulations in the battle against AI-enhanced social engineering attacks. In an environment characterized by the rapid evolution of AI models and a lack of training data, defending against generative AI threats requires constant adaptation and the collective efforts of individuals, organizations, and governments. This research seeks to provide a comprehensive understanding of the dynamic interplay between generative AI and social engineering attacks, equipping stakeholders with the knowledge to navigate this intricate cybersecurity landscape.

# I. INTRODUCTION

The evolving landscape of cybersecurity has witnessed a formidable adversary in the form of social engineering attacks. These manipulative tactics exploit human psychology and trust, often leading to unauthorized access, data breaches, and financial losses [1]. As technology continues to advance, social engineering attackers are increasingly leveraging sophisticated tools and techniques to deceive their targets. One such tool that has garnered considerable attention in recent years is generative artificial intelligence (AI) [2].

Generative AI, a subset of artificial intelligence, encompasses a range of algorithms and models capable of generating text, images, audio, and more with human-like characteristics. This technology has shown promise in various domains, including creative content generation, language translation, and medical diagnosis [3]. However, it has also found a dark niche in the world of cybercrime, where it is harnessed to enhance the efficacy and believability of social engineering attacks [4].

This research seeks to delve into the intersection of generative AI and social engineering attacks, with a specific focus on the utilization of generative AI techniques through blog content extraction. Blog extraction, a method of collecting information from publicly available blogs and online content, offers a unique perspective on understanding the emergence of AI-enhanced social engineering tactics.

# II. LITERATURE REVIEW

A This section examines pertinent literature concerning social engineering attacks and the role of generative AI within the realm of cybersecurity.

## A. Social Engineering Attacks

Social engineering attacks represent a persistent and evolving threat to cybersecurity. These attacks leverage psychological manipulation and social interaction to deceive individuals or organizations into divulging confidential information, granting unauthorized access, or performing actions that compromise security [4]. Such attacks come in various forms, including phishing, pretexting, baiting, and tailgating, among others [5]. Over the years, they have continued to adapt and remain successful due to their exploitation of human vulnerabilities [5].

Phishing attacks have been growing in frequency and complexity, posing a heightened risk to anyone who uses digital communication methods like email and text messaging. These attacks involve cybercriminals sending deceptive messages with the intention of duping recipients into divulging sensitive information, such as credit card details, or initiating malware on their devices. The effectiveness of phishing attacks has increased significantly due to the refinement of these tactics [6].

A study conducted in October 2022 by messaging security provider SlashNext revealed some concerning statistics. The research analysed billions of link-based URLs, attachments, and natural language messages across various digital communication channels over six months. It identified more than 255 million phishing attacks during this time, marking a striking 61% increase compared to the previous year [6].

These attacks often lead to financial losses, reputational damage, and data breaches, making them a focal point for cybersecurity professionals [7].

## B. Generative AI in Cybersecurity

Generative artificial intelligence (AI) is a subset of AI that encompasses various machine learning techniques capable of generating content that mimics human characteristics. Prominent among these techniques is the use of deep learning models, such as recurrent neural networks (RNNs) and generative adversarial networks (GANs), which have demonstrated remarkable capabilities in text, image, and audio generation [3].

Generative AI has found applications across diverse domains, including natural language processing, creative content generation, and even medical diagnosis. Its ability to generate content that closely resembles human-produced material has raised both opportunities and concerns in the field of cybersecurity [8].

Recent studies and anecdotal evidence have pointed to the integration of generative AI techniques into social engineering attacks. Attackers are harnessing generative AI to craft more convincing and contextually relevant phishing emails, chatbot interactions, and voice-based impersonations. These AI-enhanced attacks leverage generative AI's ability to generate persuasive content that bypasses traditional security measures [9].

## III. METHODOLOGY

This section outlines the methodology employed in this research, encompassing the data collection, data extraction, and analysis processes.

### A. Data Collection

Within this section, we elucidate the procedure for blog mining and the criteria used for selecting blogs.

This study utilizes a blog mining approach to procure data on social engineering attacks involving generative AI. Blog mining is a systematic process for extracting information from publicly available blogs and online content [10]. It is an invaluable source for threat intelligence as it provides an unfiltered perspective on the evolving trends, tactics, and techniques employed by malicious actors, shedding light on the integration of generative AI in social engineering attacks.

Blog mining has been instrumental in monitoring emerging threats, tracking the evolution of attack methodologies, and dissecting attacker motivations. This methodology grants access to social engineering incidents, offering a unique viewpoint on the intersection of generative AI and cybercrime.

Given that generative AI is a rapidly evolving field, technology consultants and security experts have frequently shared their insights and concerns via blogs. Consequently, these blogs constitute a valuable source for comprehending the intricacies associated with generative AI. However, it is essential to acknowledge the limitation of blog mining: the content is not peer-reviewed akin to journal publications and often reflects personal opinions and attitudes. To mitigate this limitation, a robust approach combines blog mining with an extensive literature search to gain a more comprehensive understanding of the topics under scrutiny.

To ensure a comprehensive data collection process, we employed Google Blog Search, which is designed to retrieve publicly available content from blogs on the internet. We recognize the changes in Google's services, and although Google Blog Search is no longer active, we conducted our search through the Google search engine. The specific keyword used for the search was "the use of generative AI in social engineering attacks." This search was conducted on September 19, 2023, resulting in 76 blogs retrieved within 0.34 seconds, predominantly from the year 2023. Subsequently, to curate the most relevant content, we selected the 'News' category and sorted the results by relevance and recency.

### B. Data Extraction and Analysis

Each of the 76 identified blogs underwent manual scrutiny to systematically extract content pertinent to this research. Inclusion criteria were defined to incorporate blogs that addressed the utilization of generative AI in social engineering attacks. Ultimately, 39 blogs met these criteria, while the remaining 37 out of the 76 were excluded for various reasons. These exclusions were attributed to factors such as the lack of focus on the use of generative AI in social engineering attacks, brief mentions of generative AI in social engineering within advertising blogs, or the blogs falling under categories like announcements, reports, news, or being blocked websites. Some blogs also necessitated subscription access. While a few blogs centred on topics closely related to our research, they did not directly address the primary research topic.

## IV. GENERATIVE AI IN SOCIAL ENGINEERING ATTACKS

The landscape of social engineering attacks has undergone substantial evolution, marked by a growing utilization of generative AI by malicious actors to enhance their strategies and heighten the prospects of success. In this section, we explore the application of generative AI in social engineering attacks, offering concrete case studies, insights into vulnerabilities exploited, and an assessment of the resultant impact.

Generative AI's fundamental aim lies in producing data that is virtually indistinguishable from authentic data, mimicking human creation or conforming to the original data's distribution. This versatility finds applications across diverse domains, including natural language generation, image synthesis, music composition, and even video generation [11].

Artificial intelligence (AI)-powered generative models possess the potential to generate exceptionally convincing deepfake content, presenting a substantial threat in the realm of malicious activities. These advanced AI systems can fabricate remarkably realistic media, virtually indistinguishable from genuine content. Consequently, they pose formidable challenges in discerning and combatting the proliferation of misinformation, manipulative social engineering attacks, and deceptive disinformation campaigns [12].

The proliferation of AI has been widely documented, but what often goes unremarked is the concurrent increase in AI's sophistication, which has introduced a fresh avenue for business email compromise attacks [13].

Generative AI encompasses any form of artificial intelligence (AI) capable of generating novel content, spanning text, music, images, code, or any other data format. Among these, OpenAI's ChatGPT celebrated as the fastest-growing consumer application in history, has garnered substantial public attention [14].

Furthermore, the automation of phishing email generation holds the potential to enable cybercriminals to produce malicious emails at an unprecedented scale, far exceeding the capacity of human attackers. AI endowed with social engineering capabilities not only can disseminate an extensive volume of initial emails but can also adapt its strategy in real-time, underscoring the multifaceted nature of this threat [14].

A recent research analysis conducted by Darktrace unveiled a startling 135% surge in social engineering attacks leveraging generative AI [15]. Cybercriminals exploit these tools for password hacking, confidential information leaks, and scams across diverse platforms. This emergence of novel scams has induced heightened apprehension among employees, with 82% expressing concerns about falling victim to these deceptive tactics [15].

### A. Generative AI Commonly Used in Social Engineering Attacks

The following are some of the commonly used generative AI tools used in social engineering attacks.

1. ChatGPT: ChatGPT is an advanced AI language model developed by OpenAI. It is part of the GPT-3

family of models and is specifically designed for natural language understanding and generation. ChatGPT is known for its ability to engage in text-based conversations with users, answer questions, provide information, and generate human-like text. It has a wide range of applications, including chatbots, virtual assistants, customer support, content generation, and more. ChatGPT's versatility and natural language capabilities make it a valuable tool in various industries and domains [16], [17].

2. FraudGPT- a game changer in malicious cyber-attacks: FraudGPT is a novel subscription-based generative AI tool designed to transcend the boundaries of technology's intended use and circumvent restrictions, opening the door to the development of highly convincing phishing emails and deceptive websites [18]. Unearthed by Netenrich's threat research team in July 2023 within the dark web's Telegram channels, FraudGPT represents a paradigm shift in attack tradecraft, potentially democratizing weaponized generative AI on a large scale [18].

Engineered to automate an array of tasks, from crafting malicious code and conceiving undetectable malware to composing persuasive phishing emails, FraudGPT empowers even inexperienced attackers. This cyber-attacker's starter kit leverages proven attack tools, encompassing custom hacking guides, vulnerability mining, and zero-day exploits, all without necessitating advanced technical expertise [19]. For a monthly fee of $200 or an annual subscription of $1,700, FraudGPT furnishes subscribers with a foundational skill set that previously demanded substantial effort to cultivate. Its capabilities span the creation of phishing emails and social engineering content, the development of exploits, malware, and hacking tools, the discovery of vulnerabilities, compromised credentials, and exploitable websites, and the provision of guidance on hacking techniques and cybercrime [19].

FraudGPT marks the onset of a new era characterized by more perilous and democratized weaponized generative AI tools and applications. While the current iteration may not match the generative AI prowess wielded by nation-state attack teams and large-scale operations, such as the North Korean Army's elite Reconnaissance General Bureau's cyberwarfare arm, Department 121, it excels in training the next generation of attackers [19].

3. WormGPT- enhancing email attacks with AI: While OpenAI LP's ChatGPT garners substantial attention in the AI landscape, hackers have embraced its "black hat" counterpart, WormGPT, to craft compelling, personalized emails that significantly elevate the success rate of their attacks. WormGPT, based on the GPTJ language model, is specifically trained to bolster the development of such malicious endeavours [13]. Initially conceived in 2021, WormGPT leverages the GPTJ language model to offer an augmented feature set, including unrestricted character support, chat memory retention, and code formatting capabilities. Unlike its ethical counterparts, WormGPT is purpose-built for nefarious activities and has exhibited the capability to generate astute and persuasive BEC (Business Email Compromise) emails [13].

4. Microsoft’s VALL-E and the emergence of voice cloning scams: On January 9th, 2023 Microsoft unveiled Vall-E, a generative AI-powered voice simulator capable of replicating a user's voice and delivering responses in the user's unique tonality, utilizing only a brief three-second audio sample [20]. Other similar tools, such as Sensory and Resemble AI, have also surfaced. Scammers have seized upon these technologies, particularly targeting Indian users, who are especially susceptible to voice-based financial scams [21].

The McAfee data underscores that up to 70% of Indian users are likely to respond to voice queries from individuals claiming to be friends or family in need of financial aid due to purported thefts, accidents, or other emergencies. This figure stands in stark contrast to users in Japan, France, Germany, and Australia, where response rates are significantly lower. Indian users, in particular, lead in sharing voice data on social media platforms, providing scammers with ample material for cloning voices through AI algorithms, and facilitating financial scams[21].

5. DALLE-E 2 and stable diffusion – the evolving landscape of AI-generated images: AI diffusion models like DALL-E 2 or Stable Diffusion are instrumental in crafting images that, while initially visibly artificial, are continually improving in quality, making it increasingly difficult to distinguish them as fake with each iteration. Concerns have arisen regarding the potential misuse of AI-generated images in disinformation campaigns and efforts to tarnish the reputation of prominent individuals, including CEOs and politicians [20].

B. Types of Social Engineering Attacks Generated by AI

1. Phishing attacks and generative AI: Phishing attacks have taken on new dimensions with the utilization of generative AI. Hackers leverage this technology to generate hyper-realistic elements, including emails, websites, and user interfaces. These sophisticated deceptions are designed to dupe individuals into divulging sensitive information or unknowingly downloading malware. As generative AI continues to advance, the complexity of these attacks is expected to increase, posing significant challenges to cybersecurity [22].

2. Pretexting and generative AI: Social engineering, a pervasive threat in the cybersecurity landscape, is experiencing notable growth, primarily driven by a technique known as pretexting. Pretexting involves the use of fabricated stories or pretexts to deceive users into divulging sensitive information [23]. Recent trends indicate a significant uptick in pretexting incidents, with these attacks accounting for nearly half of all social engineering hacks. The continued development of generative AI stands to further empower pretexting, driven by three key factors [23].

Pretexting, a form of social engineering, is on the rise and stands to benefit substantially from generative AI. AI's ability to mimic trusted organizations' writing styles enhances the credibility of pretexting attacks. Moreover, AI can significantly increase the scale of these attacks, casting wider nets across different languages, even if the threat actors lack linguistic proficiency. AI also streamlines large-scale attacks, making them more efficient. Consequently, pretexting attacks are becoming increasingly costly for organizations, necessitating vigilant cybersecurity measures [23].

3. Scams and generative AI: Generative AI poses a significant threat in the realm of scams, capable of generating fraudulent content and communications at scale to deceive individuals and organizations. These AI-driven threats compile extensive datasets containing personal and company information, using this data to craft highly unique and persuasive content, emulating writing styles and engagement patterns [24]. Scams often exploit emotions, curiosity, or urgency to deceive targets. Protecting against such scams requires robust measures, including email filters, rigorous validation of account changes, up-to-date antivirus software, and website scanners [24].

4. Deepfake social engineering: The rise of deepfake technology, fuelled by generative AI, introduces profound security challenges. Deepfakes entail the creation of highly realistic fake videos and images that convincingly mimic real people or events. These manipulated media can be used maliciously to spread misinformation, engage in impersonation, and perpetrate fraud. Deepfake threats particularly target human emotions, aiming to extract sensitive information or monetary gains. Their growing prevalence poses significant security concerns for both consumers and businesses, necessitating vigilant monitoring and protective measures [25].

Deepfakes leverage deep learning techniques, such as generative adversarial networks, to digitally alter and simulate real individuals. Instances of deepfake attacks are on the rise as the technology becomes increasingly realistic and challenging to detect. These attacks can have severe consequences, including political disinformation and financial fraud. AI tools' improvements have made it easier for perpetrators to disrupt business operations, highlighting the need for heightened awareness and robust security practices [26].

C. How AI is Advancing Social Engineering

AI is ushering in a new era in the realm of social engineering, presenting threat actors and cybercriminals with advanced tools and tactics to manipulate, deceive, and compromise computer systems. These technological advancements are creating several avenues for adversaries to employ AI in the orchestration of sophisticated social engineering attacks.

1. Enhanced phishing emails: Traditional phishing attacks often contain noticeable grammatical errors, particularly when conducted by non-native speakers. However, AI-powered tools like ChatGPT enable attackers to craft exceptionally refined emails with correct grammar and spelling, making them indistinguishable from human-authored messages. This heightened level of sophistication poses a considerable challenge in discerning AI-generated content from genuine human interactions [9]. Generative AI can produce fraudulent content and digital interactions, including real-time conversations, to impersonate users and elevate social engineering and phishing attacks. It allows non-native English speakers to refine their messages and avoid common linguistic pitfalls [27].

2. Deepfake Videos and Virtual Identities: AI can be harnessed to create convincing deepfakes, including synthetic videos and fake virtual identities. Cybercriminals can impersonate real individuals, such as senior executives or trusted partners, engaging victims in conversations to socially engineer them into revealing sensitive information, executing financial transactions, or propagating misinformation [9]. Believable deepfake videos can deceive users into taking action and divulging credentials, potentially undermining the effectiveness of employee cybersecurity training [28].

3. Voice cloning for vishing attacks: Threat actors can leverage AI to clone human speech and audio, enabling advanced voice phishing or "vishing" attacks. Scammers may use AI voice cloning technology to impersonate family members or authoritative figures, duping victims into transferring money under the pretext of a family emergency, as cautioned by the Federal Trade Commission [9]. Audio AI tools simulate the voices of managers and senior executives, enabling the creation of fraudulent voice memos or other communications with instructions for staff [28].

4. AI-driven phishing tools: AI tools can also serve as potent instruments for phishing. For example, through a complex technique called Indirect Prompt Injection, researchers successfully manipulated a Bing chatbot into impersonating a Microsoft employee and generating phishing messages that solicited credit card information from users [9].

5. Automation for industrial-scale attacks: Autonomous agents, scripting, and other automation tools empower threat actors to execute highly targeted social engineering attacks on an industrial scale. They can automate every step of the process, from selecting targets to delivering phishing emails and orchestrating human-like responses in chat boxes or phone calls [9], [29].

6. AI's adaptive learning: AI systems can adapt and refine their phishing tactics based on their own learning experiences, distinguishing between what works and what does not. This adaptive capability enables them to evolve increasingly sophisticated phishing strategies [9].

C. Vulnerabilities Exploited by Generative AI
AI-driven vulnerabilities in the realm of social engineering have introduced new dimensions of exploitation, leveraging the psychological and behavioural aspects of individuals.

1. Psychological manipulation: Generative AI equips attackers with the capability to craft highly persuasive messages that exploit human cognitive biases and emotions. By generating content that triggers recipients' fears, desires, or a sense of urgency, attackers manipulate individuals into taking actions they would not typically consider [30]. This psychological manipulation extends to various domains, including social media platforms, where phishing campaigns and deepfakes leverage social engineering to encourage victims to divulge confidential login credentials. This stolen information can then be weaponized to compromise social media accounts or infiltrate corporate networks via the infected user's device or abused credentials [30].

2. Targeted phishing: AI-driven phishing attacks have grown in sophistication, with attackers utilizing generative AI to personalize messages based on the recipient's online activities and behaviours. These targeted phishing attempts are particularly challenging to detect as they often emulate trusted sources and employ social engineering tactics tailored to the specific victim. Generative AI enables attackers to craft convincing phishing lures with personalization, drawing from information available on public profiles like LinkedIn. This personalization amplifies the effectiveness of business email compromise (BEC) attacks by generating emails that align closely with the target's context, language, and tone [12], [31].

3. Crisis of authenticity: Generative AI threatens to produce highly credible content that emulates individual or organizational writing styles. These materials are designed to resonate emotionally with consumers, exploiting their trust and emotions. Attackers leverage deepfake technology in outbound and inbound voice or video calls, relying on victims' emotions, panic, and curiosity to establish trust, urgency, and empathy [24]. Generative AI models consider various data points to achieve their objectives, often involving actions such as sharing sensitive information or transferring funds to fraudulent accounts. The resulting conversations sound authentic and empathetic, making it essential for individuals to remain vigilant, scrutinize details, and contact their banks' fraud hotlines if they suspect fraudulent activity [24].

Generative AI technology raises concerns about authenticity and governance in digital media. It poses immediate risks of disinformation, potentially leading to longer-term issues related to control and accountability. AI-powered bots can flood social media platforms with tailored content, spreading rapidly and amplifying contentious issues. Businesses will need to monitor digital media manipulation carefully, as generative AI enables the creation of hyper-targeted, AI-generated content at scale. This capability elevates the risk of campaigns related to "augmented advocacy," activist boycotts, and online backlash surrounding sensitive topics [32].

4. Deepfake technology: Deepfake technology, a subset of generative AI, empowers attackers to produce highly convincing audio and video impersonations. These deepfake voice recordings or video clips allow attackers to impersonate trusted individuals or authority figures, eroding trust and facilitating manipulation. Deepfake technologies also introduce fabricated facial overlays during video calls, creating a false image and voice on the consumer's screen. This innovation is akin to wearing a mask and mimicking a voice during interactions, posing significant risks in various contexts [24].

D. Impact and Consequences

The misuse of AI-powered generative models carries profound and far-reaching consequences, with the potential to disrupt trust, manipulate public opinion, and inflict substantial harm on individuals, institutions, and even entire societies. Vigilance and the development of robust countermeasures are imperative to mitigate the risks associated with this evolving technology [12].

1. Financial losses: AI-enhanced social engineering attacks often result in significant financial losses for both individuals and organizations. Victims may unwittingly transfer funds, divulge sensitive financial information, or engage in transactions that directly benefit the attackers. Such monetary losses can have severe repercussions [9].

2. Reputation damage: Social engineering attacks can tarnish the reputation of targeted individuals and organizations. When AI-generated content is exploited to disseminate false information or conduct smear campaigns, the harm to reputation can be enduring and challenging to remediate. This damage extends beyond the financial realm, affecting trust and credibility [7].

Deepfake technology introduces the unsettling possibility of highly damaging revenge scenarios. Individuals with malicious intent could easily fabricate scenarios such as making it appear as though someone cheated by swapping faces in an intimate video. They could create fake videos depicting the victim making offensive statements, thereby jeopardizing their career, even if the video is later proven to be fake. Blackmailing someone with a deepfake video, and threatening to release it publicly unless certain demands are met, is another alarming prospect [33].

3. Legal Implications: The utilization of generative AI in social engineering attacks gives rise to intricate legal challenges, particularly concerning issues of attribution and accountability. As AI technology continues to advance, identifying the true identities of attackers and holding them responsible becomes increasingly complex and, in some cases, elusive [34].

These consequences underscore the critical need for comprehensive strategies to combat the misuse of AI-powered generative models. Implementing cybersecurity measures, raising public awareness, and fostering responsible AI development are crucial steps toward addressing the multifaceted risks associated with this emerging technology.

E. Real-World Examples of AI-Enhanced Social Engineering Attacks

Through blog mining, we have identified numerous instances where generative AI techniques were employed in social engineering attacks. These case studies provide concrete examples of how attackers leverage AI in their efforts. AI-powered generative models have been involved in various incidents that demonstrate the potential for misuse and deception.

1. Tricking ChatGPT for Windows activation keys: Users were able to deceive ChatGPT into providing Windows activation keys by framing their request as part of a bedtime story [35].

2. Requesting a deceptive email: A user requested ChatGPT to compose a deceptive email, posing as a friendly yet professional message regarding an account issue and instructing the recipient to call a specific phone number [36].

3. Impersonation of UK business owner: In 2020, AI technology was used to impersonate the voice of a UK

business owner. This impersonation convinced a CEO to transfer $243,000 to an unknown group of hackers [20].

4. Voice mimicry in a UK energy company: In 2019, hackers targeted a UK energy company using AI-generated voice technology. They successfully mimicked the CEO's voice and persuaded a subordinate to transfer approximately $243,000 to a fraudulent account [27].

5. Blackmail using AI-generated voice: Criminals blackmailed a woman by claiming they had kidnapped her daughter and used AI to create a convincing simulation using voice samples from the daughter. In February 2023, a journalist broke past the authentication scheme of a major UK financial institution by using deepfake technology [37].

6. $35 Million Stolen in UAE: A criminal ring stole $35 million by using deepfake audio to deceive an employee at a United Arab Emirates company. They convinced the employee that a director needed the money for an acquisition on behalf of the organization [38].

7. Impersonation of CEO: Scammers impersonated a CEO using deepfake audio in an attempt to request a transfer of €220,000 from a manager of a U.K. subsidiary of a German company [34].

8. Face-swapping scam in China: In northern China, a cybercriminal used AI-powered face-swapping technology to impersonate a man's close friend, persuading him to transfer 4.3 million yuan [34].

These incidents highlight the capacity of AI-generated content and voice technology to facilitate deception, social engineering, and financial fraud. As AI tools become more sophisticated, there is a growing need for awareness and countermeasures to protect against such threats.

### F. Challenges in Detecting AI-Enhanced Attacks

The challenges associated with detecting AI-enhanced attacks, particularly those involving generative AI, are multifaceted and require specialized approaches:

1. Evolving AI models: The rapid evolution of generative AI models poses a significant challenge to defenders. Attackers can quickly adapt their tactics to bypass existing detection methods by leveraging the latest AI techniques. This constant evolution demands continuous vigilance and proactive measures to stay ahead of emerging threats. Generative AI models are often considered black boxes, making it challenging to understand their decision-making processes. This opacity complicates the attribution of responsibility, hindering effective countermeasures and legal actions [22].

2. Lack of training data: Detecting AI-enhanced social engineering attacks relies on large and diverse datasets for training machine learning models. However, obtaining labelled datasets that cover the full spectrum of potential attacks can be a daunting task. Generative AI can be used by cybercriminals to automate social engineering attacks, such as highly personalized spear-phishing campaigns. By analysing vast amounts of data, AI can craft convincing messages targeting specific individuals or groups, thereby increasing the success rates for malicious actors. Overcoming these challenges necessitates access to extensive training data, which can be difficult to obtain due to privacy concerns and legal constraints [22].

To effectively combat and mitigate generative AI threats, the cybersecurity community must address these challenges through ongoing research, collaboration, and the development of advanced detection techniques. Staying updated on AI advancements and continuously improving defence strategies are essential components of defending against AI-enhanced attacks. Additionally, addressing the privacy concerns associated with data collection and sharing is crucial to strike a balance between security and individual rights.

G. Countermeasures and Defence Against AI-Enhanced Social Engineering Attacks

Addressing the growing threat of AI-enhanced social engineering attacks requires a multi-faceted approach that combines traditional security measures with advanced AI-based solutions.

1. Traditional Security Measures: Continue to employ traditional cybersecurity practices, such as firewalls, intrusion detection systems, and email filtering. These mechanisms help detect known attack patterns and malicious entities [35].

2. Advanced Email Filters and Antivirus Software: Enhance your email security with advanced filters that can identify phishing attempts and malicious content. Keep antivirus software updated and relevant to protect against malware [39].

3. Website Scanners: Utilize website scanning services to identify false or malicious websites. These scanners can help users avoid falling victim to phishing sites [25].

4. Multi-Factor Authentication (MFA): Implement MFA, especially for sensitive transactions or approvals. MFA adds an extra layer of security by requiring multiple forms of verification [27], [35].

5. Phishing Simulations: Conduct phishing simulations using AI tools like ChatGPT to familiarize employees with the tone and characteristics of AI-generated communications. This can help employees recognize and respond to phishing attempts effectively[36].

6. Implement Passwordless Authentication: Consider using passwordless authentication systems like passkeys that use cryptography to make user credentials unphishable [18]. These systems can significantly enhance security by eliminating the reliance on traditional passwords. Embrace phishing-resistant factors like passkeys, which can help combat advanced phishing attempts facilitated by AI [40] [18].

7. AI-Powered Security Solutions: Invest in AI-driven security solutions that leverage machine learning algorithms to analyse network traffic, email content, and user behaviour for anomalies indicative of social engineering attacks. Fighting AI with AI [28], [35], [36], [41]–[43].

8. Enhance AI-Driven Threat Detection: Develop machine learning models capable of recognizing subtle AI-generated content patterns and behaviours. Integrating AI with traditional security measures creates a more comprehensive defence [31].

9. Collaborative approaches: Foster collaboration among cybersecurity professionals, organizations, and AI developers. Sharing threat intelligence and forming public-private partnerships can enhance defence strategies [22].

10. Zero trust framework: Embrace a zero-trust approach, which assumes that no one, whether inside or outside the organization, should be trusted by default. Verify the senders of emails, chats, or texts, especially when they request sensitive information or actions [23], [32], [36].

11. Awareness and Education: Educate employees and individuals about the risks associated with generative AI and social engineering attacks. Promote awareness of the evolving threat landscape [14], [27], [32], [36], [44].

12. Continuous improvement: Recognize that the threat landscape is constantly evolving. Stay agile and adapt defence strategies to counter new and emerging threats effectively. Regularly assess and improve security measures. Keep pace with advancements in AI technology and evolving attack techniques [12], [28].
By combining these strategies and maintaining a proactive stance, individuals, organizations, and governments can work together to reduce the risk of AI misuse by cybercriminals and create a safer digital environment for all.

The battle against AI-enhanced social engineering attacks is ongoing, and as attackers continue to exploit AI technology, defenders must adapt their strategies to stay one step ahead. As organizations strive to combat generative AI threats, they must navigate the delicate balance between security measures and privacy concerns. Mitigation efforts should avoid unnecessary invasions of privacy while still protecting individuals and organizations from potential harm.

## V. CONCLUSION AND FURTHER RESEARCH

In this study, we explore the relationship between artificial intelligence (AI) and social engineering attacks, using a research approach known as blog mining. We investigate real-world cases, analyse how AI plays a role in these attacks, examine their consequences, and assess the strategies employed for defence.

One key finding from our research is the growing utilization of AI by malicious actors to enhance the effectiveness of their social engineering tactics, a development that raises significant concerns about its potential misuse.

Furthermore, our study highlights the increasing complexity and success rate of AI-backed social engineering attacks. Attackers can now craft highly personalized and contextually relevant messages that pose significant challenges for detection.

Conventional cybersecurity measures face difficulties in effectively countering AI-enhanced attacks due to the rapid evolution of AI models and the extensive data required for training. As a result, there is a pressing need for adaptation and innovation in the cybersecurity landscape.

Looking ahead, several critical areas warrant further exploration:

The development of AI-driven threat detection systems capable of identifying subtle patterns and behaviours indicative of AI-generated threats should be prioritized. These systems should complement existing security measures.

Collaboration between various sectors, including public and private entities, is essential for sharing threat intelligence, anticipating emerging threats, and developing proactive defence strategies.

Emphasizing ethical considerations, transparency, fairness, and accountability in the development and deployment of AI technologies can help mitigate the risks associated with AI-enhanced social engineering attacks.

In conclusion, the integration of generative AI into social engineering attacks presents a formidable challenge for cybersecurity. While AI holds significant promise, its potential for misuse necessitates vigilance, ethical guidelines, and legal frameworks. Addressing these multifaceted challenges requires a collaborative effort involving technology developers, organizations, policymakers, and society at large to strike a balance between innovation and security.